\documentclass[]{spie}  
\usepackage{url}            
\usepackage[]{graphicx}
\usepackage[]{graphicx}

\title{VADER - A Satellite Mission Concept For High Precision Dark Energy Studies} 

\author{Rene Fassbender\supit{a}, Jutta Stegmaier\supit{b}, Anne-Marie Weijmans\supit{c}, Stefan K\"ostner\supit{d}, \\
Andreas Kruselburger\supit{e}, Cornelia Diethart\supit{f}, Peter Fertl\supit{g}, Elisabetta Valiante\supit{a},  \\
Matthew Hayes\supit{h}, Peter Schuecker\supit{a} and G\"unther Hasinger\supit{a}
\skiplinehalf
\supit{a}Max-Planck-Institut f\"ur extraterrestrische Physik, Giessenbachstrasse 1, D-85748 Garching, Germany; \\
\supit{b}Max-Planck-Institut f\"ur Astronomie, K\"onigsstuhl 17, D-69117 Heidelberg, Germany; \\
\supit{c}Leiden Observatory, P.O. Box 9513, NL-2300 Leiden, The Netherlands; \\
\supit{d}Atominstitut der \"Osterreichischen Universit\"aten, Stadionallee 2, A-1020 Wien, Austria; \\
\supit{e}Technische Universit\"at M\"unchen, Institute of Astronautics, Boltzmannstrasse 15, D-85748 Garching, Germany; \\
\supit{f}Institut f\"ur Astronomie der Universit\"at Wien, T\"urkenschanzstrasse 17, A-1180 Wien, Austria; \\
\supit{g}Technische Universit\"at Wien, Institute of Communications and Radio Frequency Engineering, Gusshausstrasse 25/389, A-1040 Wien, Austria; \\
\supit{h} University of Stockholm, Albanova University Center, SE-10691 Stockholm, Sweden
}

\authorinfo{Further author information: (Send correspondence to Rene Fassbender)\\R. Fassbender: E-mail: rfassben@mpe.mpg.de, Telephone: +49 89 30000 3319}

\def\sun{\hbox{$\odot$}}                                    
\def\arcmin{\hbox{$^\prime$}}                               
\def\arcsec{\hbox{$^{\prime\prime}$}}                       
\def\degr{\hbox{$^\circ$}}                                  

\def\la{\mathrel{\mathchoice {\vcenter{\offinterlineskip\halign{\hfil
$\displaystyle##$\hfil\cr<\cr\sim\cr}}}
{\vcenter{\offinterlineskip\halign{\hfil$\textstyle##$\hfil\cr
<\cr\sim\cr}}}
{\vcenter{\offinterlineskip\halign{\hfil$\scriptstyle##$\hfil\cr
<\cr\sim\cr}}}
{\vcenter{\offinterlineskip\halign{\hfil$\scriptscriptstyle##$\hfil\cr
<\cr\sim\cr}}}}}                                            
\def\num{\hbox{N$^{\underline{o}}$}}                        

\begin{document} 
  \maketitle 

\begin{abstract}
We present a satellite mission concept to measure the dark energy equation of state parameter $w$ with percent-level precision. The Very Ambitious Dark Energy Research satellite (VADER) is a multi-wavelength survey mission joining X-ray, optical, and IR instruments for a simultaneous spectral coverage from 4\,$\mu$m (0.3\,eV) to 10\,keV over a field of view (FoV) of 1 square degree. VADER combines several clean methods for dark energy studies, the baryonic acoustic oscillations in the galaxy and galaxy cluster power spectrum and weak lensing, for a joint analysis over an unrivalled survey volume.

The payload consists of two XMM-like X-ray telescopes with an effective area of 2,800\,cm$^2$ at 1.5\,keV and  state-of-the-art wide field DEPFET pixel detectors (0.1-10\,keV) in a curved focal plane configuration to extend the FoV. The X-ray telescopes are complemented by a 1.5\,m optical/IR telescope with 8 instruments for simultaneous coverage of the same FoV from 0.3\,$\mu$m to 4\,$\mu$m. The 8 dichroic-separated  bands (u,g,r,z,J,H,K,L) provide accurate photometric galaxy redshifts, whereas the diffraction-limited resolution of the central z-band allows precise shape measurements for cosmic shear analysis. 

The 5 year VADER survey will cover a contiguous sky area of 3,500 square degrees to a depth of $z$$\sim$2 and will yield accurate photometric redshifts and multi-wavelength object parameters for about 175,000 galaxy clusters, one billion galaxies, and 5 million AGN. VADER will not only provide unprecedented constraints on the nature of dark energy, but will additionally extend and trigger a multitude of cosmic evolution studies to very large ($>$10 Gyrs) look-back times.
\end{abstract}



\section{INTRODUCTION}
\label{sect:intro}  


The nature of dark energy (DE), as the driving force of the observed acceleration of the Universe, is currently one of the deepest mysteries in astrophysics. Whereas a physically motivated theory of DE is still out of reach, upcoming observational data sets will be able to put improved constraints on the DE equation of state parameter $w$, which relates the fundamental properties pressure and energy density of DE. A precise knowledge of $w$ and in particular its redshift dependance, usually parameterized as $w(z)=w_0+w_1 \cdot z$, is required to allow at least a phenomenological understanding of the dominant energy contribution in the Universe. The answer to the fundamental question whether DE is equivalent to Einstein's cosmological constant ($w$=-1), phantom energy\cite{Caldwell2003a} ($w$$<$-1), or time varying as in quintessence models\cite{Wetterich1988a} (-1$<$$w$$<$-1/3) will not only shed light on the fate of the Universe but will also lead the way to a unified physical theory.

Current observational data can constrain $w_0$ to about 5-10\%\cite{Schuecker2005a}, and are in agreement with the value of the cosmological constant.    
A first generation of designated dark energy surveys (e.g. KIDS, APEX-SZ) is currently upcoming or in the final planning stages. Second generation ground-based experiments (HETDEX, DES, Pan-STARRS, LSST) and planned satellite missions (eROSITA, SNAP, DUNE) are foreseen to yield valuable results after the year 2010.
However, constraining $w$ with high precision will be a very challenging observational task. A final distinction between DE models, in particular with $w_0$ being close to -1, is likely to require sub-percent precisions\cite{Caldwell2005a} that are beyond the capabilities of the currently planned experiments mentioned above. Our proposed VADER survey will bridge this gap. The VADER project originated from a case study of  future DE space missions and is mainly targeted at the feasibility of the experiment without being constrained by a specific mission framework.

The paper is organized as follows. In Sect.\,2 we describe the scientific goals and the requirements for the mission. The survey design, sensitivities, and expected results are discussed in Sect.\,3. The scientific payload is presented in Sect.\,4 and the spacecraft in Sect.\,5. The conclusions are given in Sect.\,6. 
We assume concordance model cosmological parameters with $\Omega_m=0.3$, $\Omega_{\Lambda}=0.7$ and $h=0.7$, unless otherwise stated.
 

\section{SCIENCE CASE} 



\subsection{Constraining Dark Energy}

The fundamental observables\cite{Weinberg2005a} for precise measurements of the DE equation of state parameter $w$ are the Hubble parameter $H(z)$, the distance-redshift relation $d(z)$, and the linear growth factor $D_1(z)$. For example, $w$ influences the cosmic expansion history $H(z)$ for a flat Universe in the following way:

\vspace{-1ex}
\begin{equation}
\vspace{-1ex}
\label{equ_hubbleexpansion}
H(z)=H_0\sqrt{\Omega_m (1+z)^3 + \Omega_{\Lambda} \cdot \exp{\left (3 \int \limits_{0}^{z} \frac{1+w(z)}{(1+z)} dz \right)}}
\end{equation}

\noindent
Here $H_0$ denotes the present day Hubble constant, $z$ the redshift, $\Omega_m$ the total matter density in units of the critical density, and $\Omega_{\Lambda}$  the dark energy density. Since VADER will be a third generation DE experiment, we assume that the main cosmological parameters (e.g. $H_0$, $\Omega_m$, $\Omega_{\Lambda}$, $\Omega_{b}$) will be accurately known by mission start, and thus focus the discussion on the determination of $w(z)$.


\noindent
VADER's science drivers are three independent methods to constrain dark energy: \\ 
\noindent
1) {\em Baryonic Acoustic Oscillations (BAOs):} The BAOs are an imprint of the sound horizon scale at the epoch of decoupling in the matter power spectrum (PS). The acoustic horizon acts as a comoving standard ruler, whose physical size is currently determined to better than 2\% as $r_s$=147.8\,Mpc\cite{Spergel2006a}; sub-percent precision measurements are expected from the Planck satellite. The analysis of the spectral position of the baryonic `wiggles' with an amplitude of $<$5\% in the matter power spectrum  yields a purely geometrical test based on absolute distances. It is therefore a very `clean' method with low systematic errors\cite{Seo2003a}. The BAOs are sensitive to $w$ since the transformation of redshifts to comoving distances requires $H(z)$, while transforming angular separations involves the angular diameter distance $d_A(z)$. The details for using this method as a precision test have been recently worked out\cite{Koehler2006a} and can be applied to any object class. VADER will use this method in a threefold way for (i) clusters of galaxies, which are highly biased  objects (i.e.\,small numbers needed) mostly governed by simple linear gravitational physics, (ii) galaxies, as the most abundant objects, and (iii) AGN, which allow studies out to very high redshifts.\\
\noindent
2) {\em Cosmic Shear:} Cosmic shear is the gravitational lensing effect of the large-scale structure. It allows the study of the cosmic dark matter (DM) distribution and its evolution without any reference to luminous tracers and  thus is a powerful independent cosmological test with low systematics. The $w$ sensitivity enters through the distance-redshift relation $d(z)$.\\
\noindent
3) {\em Number density evolution of galaxy clusters:} 
Clusters of galaxies are very sensitive tracers of the underlying DM large-scale structure. Their number density evolution is directly related to the structure growth function $D_1(z)$ and the evolution of comoving volume elements $dV(z)$. Since the cluster mass enters in this method, a precise calibration of the mass-observable relation on the percent level is required making this cosmological test somewhat more prone to systematic effects. However, self-calibration schemes for large cluster samples have been proposed\cite{Mohr2004a} thus lowering the necessary mass calibration accuracy at the expense of some precision on the final parameter determination.





For precision measurements of $w_0$, and in particular the time dependant part $w_1$, observations out to high redshifts are a very important requirement, which are not easily fulfilled by ground-based imaging surveys. Although the DE density  $\Omega_{\Lambda}$ is probably dynamically unimportant at $z$$>$2, an extended redshift coverage is critical for numerous reasons: 
(i) pivot points at different redshifts are needed to break the intrinsic degeneracies in $w_0$ and $w_1$\cite{Schuecker2005a},
(ii) nonlinear structure growth erases the BAOs at high wave numbers k\cite{Angulo2005a}, thus high $z$ measurements are 'cleaner', easier, and can be extended to higher wiggle order, 
(iii) the cosmic volume probed increases,  which is critical for precise BAO tests, 
(iv) $w$ enters in $H(z)$ and $d(z)$ in the integral (see Eq.\,\ref{equ_hubbleexpansion}) and is thus a cumulative effect,
(v) increased redshift leverage boosts the $w$ effects on structure growth, i.e. the cluster number density, 
and (vi) many theories predict `early dark energy'\cite{Hebecker2001a} with observable effects at high $z$.



\subsection{Scientific Goals}
Being a `third generation' dark energy experiment, VADER's scientific goals and demands are very high.  
As the main science drivers for the mission, VADER has to provide: (i) several independent DE tests with low systematics, (ii) 
sub-percent constraints on the time independent part of $w$ , (iii) precise measurements of the time variation of $w$, and (iv) tests of alternative scenarios to DE (e.g.\ extra dimensions\cite{Rhodes2003a}).
The data should be of sufficient quality to allow non-parametric tests, i.e. observations are directly compared to model predictions from future hydrodynamical simulations\cite{Springel2005a}.




Table\,\ref{t_methods} lists the five dark energy tests in order of priority, their redshift range, and the instrumental requirements. The baryonic acoustic oscillations in the galaxy cluster power spectrum were identified as the cleanest test with very low systematic effects, followed by cosmic shear measurements, the galaxy BAOs, galaxy cluster number counts, and BAOs in the AGN power spectrum.

However, the theoretical and observational developments in dark energy research over the next 15 years are difficult to foresee, which implies that alternative science goals should be considered. It could be, for example, that an undetected extra dimension exactly mimics the effects of dark energy, meaning that a different geometry of the Universe drives the accelerated expansion instead of its energy content. Observational effects of extra dimensions on the large-scale power spectrum have been worked out\cite{Rhodes2003a}. Designed to conduct high precision measurements of the power spectra of galaxy clusters, galaxies, and AGN, VADER can also efficiently probe the presence of extra dimensions, in particular with the sensitivities reached during a 10\,year extended mission lifetime.

\begin{table}[t]
\begin{center}
\begin{tabular}{|c|c|c|c|c|c|c|c|}
\hline

 {\bf ID} &  {\bf Method} &  {\bf Cosm. Test} &  {\bf Test Objects} &  {\bf $z$ Range} &  {\bf X-ray} &  {\bf Opt/IR} &  {\bf Hi Res} \\

\hline\hline

1 & BAOs in galaxy cluster PS  & $H(z)$, $d_A(z)$ & galaxy clusters & 0-2  & $\times$  &  $\times$ &     \\
2 & cosmic shear  & $d(z)$  & galaxies & 0-2  &   & $\times$  &  $\times$   \\
3 & BAOs in galaxy PS  &  $H(z)$, $d_A(z)$ & galaxies  & 0-3  &   &  $\times$ &     \\
4 & cluster number counts  &  $D_1(z)$, $dV(z)$  & galaxy clusters & 0-2  & $\times$  & $\times$ &     \\
5 & BAOs in AGN PS &  H(z), $d_A(z)$  & AGN & 0-4  & $\times$  &  $\times$ &    \\

\hline
\end{tabular}
\caption[Main methods and cosmological tests]{Main methods and cosmological tests conducted with VADER. From left to right the table states the scientific priority, the method, the fundamental cosmological test, the objects, and the covered redshift range. The last three columns specify the data requirements for the given method: (i) X-ray coverage, (ii) multi-band optical/IR coverage, and (iii) diffraction limited high spatial resolution.}

\label{t_methods}
\end{center}
\end{table}

\subsection{Design Specifications}
\label{sec_specifications}
\enlargethispage{2ex}
The instrumental requirements for VADER follow from the science goals and the identified cosmological tests and are partially indicated  in the last three columns of Tab.\,\ref{t_methods}.
All methods require accurate photometric redshifts out to a minimum redshift of $z$=2, which implies deep coverage in multiple optical and IR bands. 
The cosmic shear signal is extracted from precise shape measurements of galaxies, demanding high optical quality over the whole field of view and diffraction limited resolution in at least one band.  
The galaxy cluster and AGN tests depend critically on sensitive X-ray observations for identifications and characterizations out to high redshift, whereas the resolution has to be sufficient to separate  point-like AGN from extended galaxy cluster sources. On the other hand, the  volume probed is to be maximized thus requiring wide-field capabilities of all instruments for long pointed observations within a contiguous sky area.



\section{SURVEY DESIGN}

\subsection{Mission Overview}

In order to fulfill all specifications identified in Sect.\,\ref{sec_specifications}, VADER is designed as a multi-wavelength mission covering the spectral range from infrared to X-ray (see Sect.\,\ref{sec_payload} for details). Two high-throughput XMM-Newton type X-ray telescopes (0.1-10\,keV) are complemented by a 1.5\,m wide-field optical telescope. The optical/IR beam is dichroic-separated into eight bands (ugrzJHKL) allowing simultaneous coverage from 0.3--4\,$\mu$m. The two X-ray and eight optical/IR imaging cameras all have a field of view of 1\,square degree and survey the same patch of sky (see right panel of Fig.\,\ref{rf_fig_xperformance}).

\begin{figure}[h]
\centering
\includegraphics[width=4.2cm]{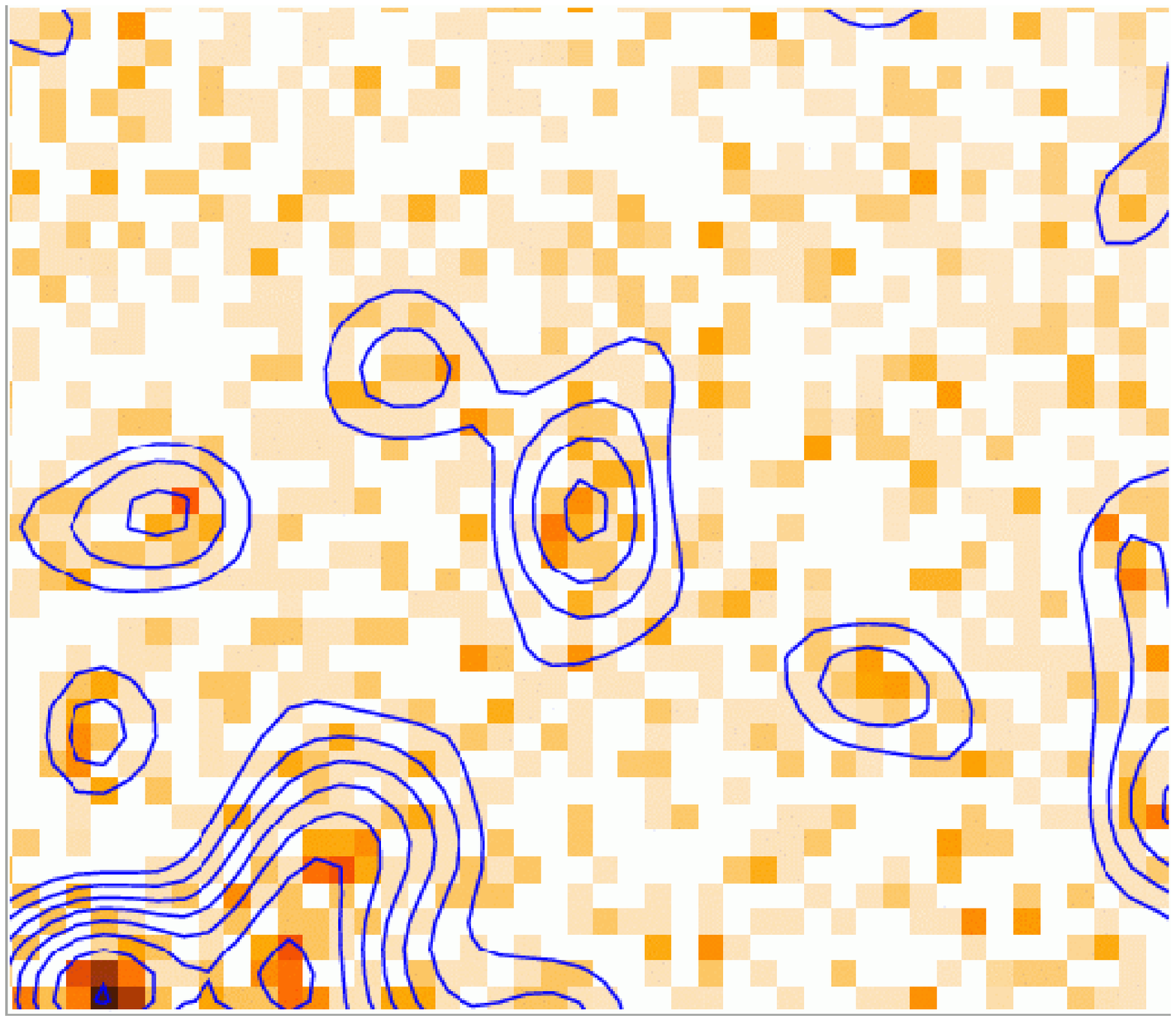}
\includegraphics[width=4.2cm]{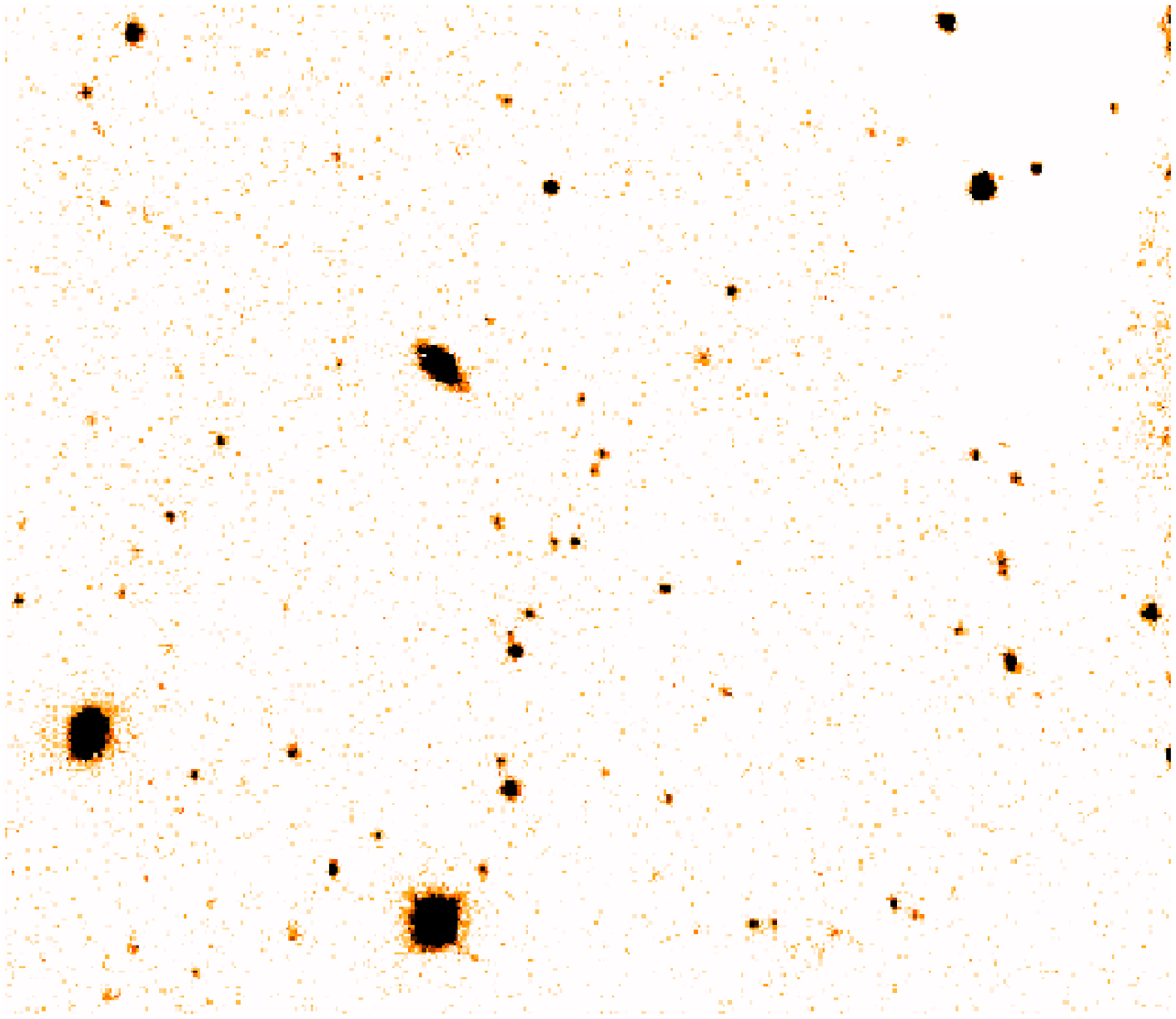}
\includegraphics[width=4.2cm]{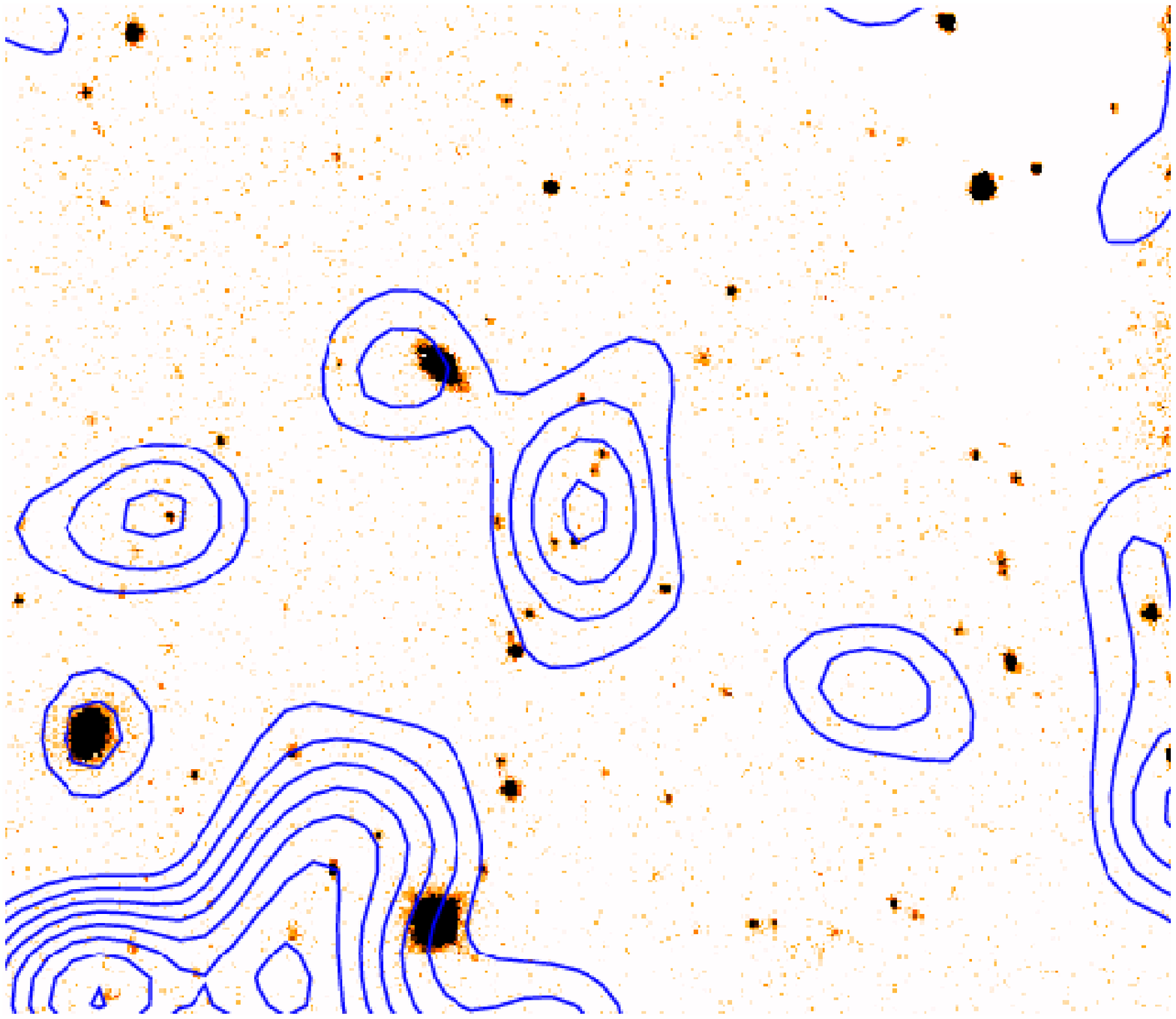}
\includegraphics[width=4.2cm]{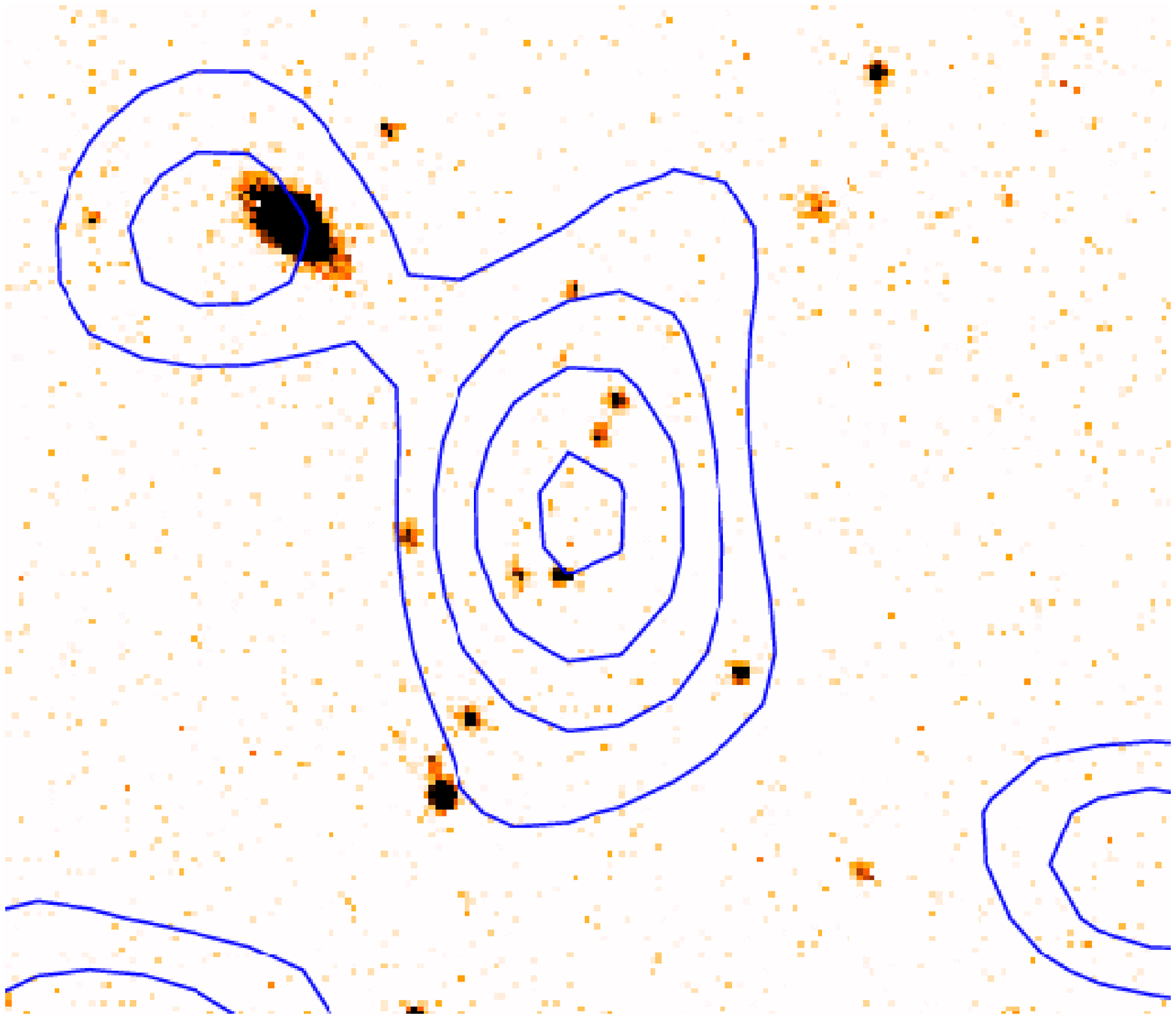}
\caption{Illustration of the sensitivity gain of the multi-wavelength source identification approach: {\bf Left:} XMM-Newton 2.5$^\prime$\,$\times$\,2.5$^\prime$ X-ray image cutout with a weak source in the center, which is well below the threshold for a reliable extent fit, i.e.\,identification as a galaxy cluster candidate is not possible. {\bf Center left:} Deep H-band image of the same sky region. The identification of a structure with optical/NIR data alone will be very difficult. {\bf Center right:} Overlay of the X-ray contours and the H-band image. The spatial coincidence of the galaxy overdensity and the X-ray emission is now obvious and allows a source identification as a potential X-ray luminous group of galaxies. {\bf Right:} 1.25$^\prime$\,$\times$\,1.25$^\prime$ zoom-in on the source. With the capabilities of VADER, the redshift, the galaxy membership, and a mass estimate of the group will be instantly available.}
\label{rf_fig_NIR_X}       
\end{figure}

The multi-wavelength approach with combined X-ray and optical/IR data is VADER's key feature and allows a substantial sensitivity gain (see Fig.\,\ref{rf_fig_NIR_X}). The matching of X-ray photons with optical data\cite{Schuecker2004a} can lower the X-ray flux limit by a factor 5-10, for VADER we conservatively assume a sensitivity gain of a factor 3 compared to pure X-ray identifications. 

The eight optical/IR bands ugrzJHKL are critical to derive accurate photometric redshifts for individual objects out to high redshifts. We aim at a redshift error of $\Delta$$z$=0.02(1+z) for individual galaxies out to $z$=3. This will allow the determination of a mean galaxy cluster redshift with $\Delta$$z$$\sim$0.01 at $z$=2 and $\Delta$$z$$<$0.01 for lower redshifts. 
Furthermore, the giga-pixel z-band camera will yield diffraction limited resolution of 0.13\arcsec \ for detailed cosmic shear tomography studies over the full survey volume.

The satellite will operate in a highly elliptical  72\,hour orbit (HEO) (see Fig.\,\ref{rf_fig_surveyregion}), which will allow 60\,h of consecutive science data acquisition outside the radiation belts.
The orbit was selected to fulfill the following requirements in an optimized way: (i) low X-ray, optical, and IR background, (ii) high science data fraction per orbit, (iii) long, pointed observations, (iv) observations of a single, contiguous survey field, (v) stable thermal environment, and (vi) high downlink rate for data transmission\footnote{For Low Earth Orbits items (iii), (iv), (v) are difficult or impossible to fulfill; L2 does not meet (vi).}.

For a nominal mission duration of 5 years, VADER will conduct a multi-wavelength survey over 3,500 square degrees, which could be increased up to 7,000 square degrees with an extended mission lifetime of 10 years.






\subsection{Survey Region \& Strategy}

\begin{figure}[t]
\centering
\includegraphics[width=7.0cm]{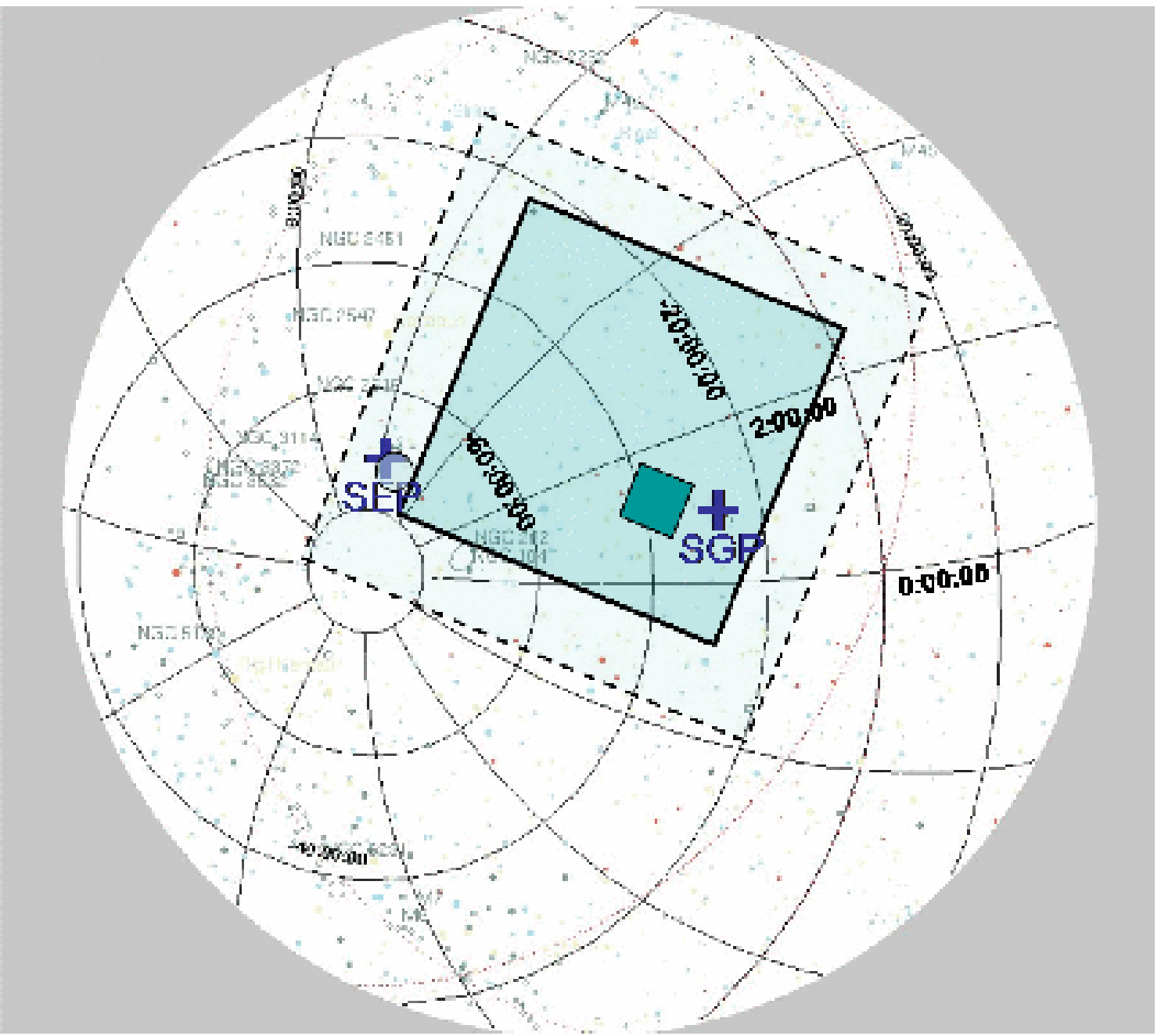}
\includegraphics[width=10.0cm]{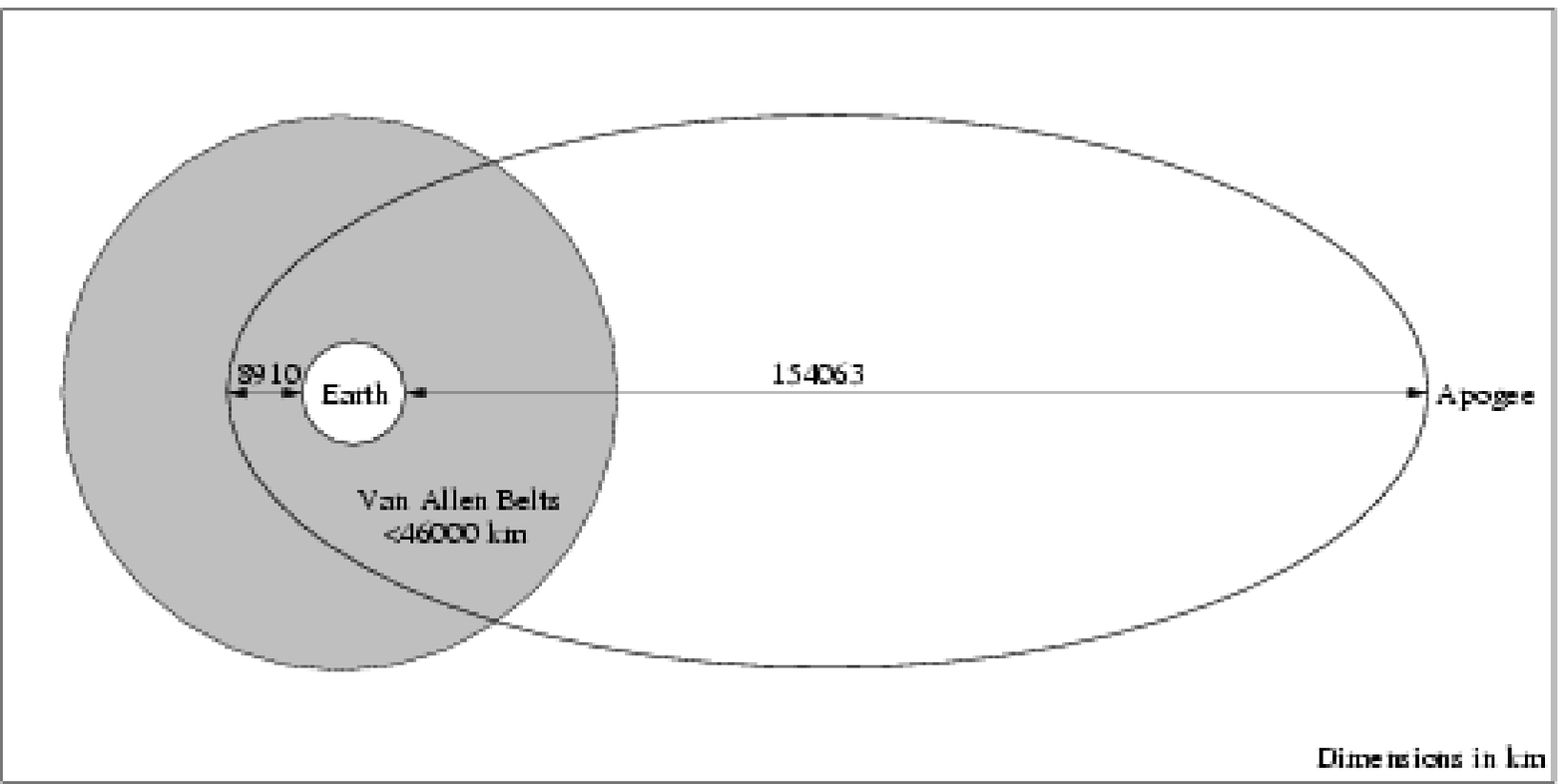}
\caption{ VADER survey region and orbit. {\bf Left:} The solid square indicates the covered sky area of 3,500\,deg$^2$ of a 5 year survey centered at RA=2.5h, DEC=-40d; the dashed one shows the prospects for an extended 10 year survey. The locations of the 100deg$^2$ deep survey field (small square), the South Galactic Pole (SGP), and the South Ecliptic Pole (SEP) are also displayed. {\bf Right:} Sketch of the 72\,h highly elliptical orbit with the radiation belt shown as shaded region. The orbit parameters are: perigee/apogee altitude: 8.910\,km/154,063\,km, semi-major axis: 87,865\,km, eccentricity: 0.83, inclination: 5\,degrees, orbital time above 50,000\,km: 63.5\,h.}
\label{rf_fig_surveyregion}       
\end{figure}

The VADER survey region is centered on RA=2.5h, DEC=-40d covering a square sky region with sides of $\sim$60 degrees for the nominal 5 year mission lifetime (see Fig.\,\ref{rf_fig_surveyregion}, left). The `cubic cone' survey volume is ideal for Fourier mode decomposition required for many of the DE tests (Tab.\,\ref{t_methods}). In addition, the region is optimized to high galactic latitudes, where the extinction and hydrogen column density are low, and to continuous observability, which is easiest at the ecliptic pole.

The satellite spends a period of 63.5\,h per orbit at altitudes above 50,000\,km. Subtracting calibration time and some extra margins, 60\,h or 83\% of the orbit period can be used for scientific data acquisition. Aiming for a gross exposure time of 10\,h per field, VADER can survey 6 square degrees per orbit or 2 square degrees per day down to the full flux/magnitide limit. This way, an annual survey coverage of 700 square degrees can be achieved, including a smaller deep field (see Fig.\,\ref{rf_fig_surveyregion}) with 30h gross exposure time that will be important to check systematics, selection functions, and perform studies at lower flux limits.

In order to allow time domain astrophysics, e.g. for very distant supernova searches, VADER will observe a field for 2\,h consecutively and then move to the next neighboring field. Every field is re-observed  four more times during the following orbits with spatial offsets\footnote{See for example the COSMOS survey strategy: \url{http://www.mpe.mpg.de/xmmcosmos/Cosmos}} between the pointings to guarantee a homogenous coverage, in particular for the X-ray survey. Within a few weeks after the first observation, the full 10h effective survey depths is built up for every individual pointing. This will allow deep monitoring of about 20 fields for time variations on day scales down to limiting magnitudes  about 1 lower than stated in Tab.\,\ref{t_fluxlimits}.




\subsection{Sensitivities}

\begin{figure}[t]
\centering
\includegraphics[width=8.0cm]{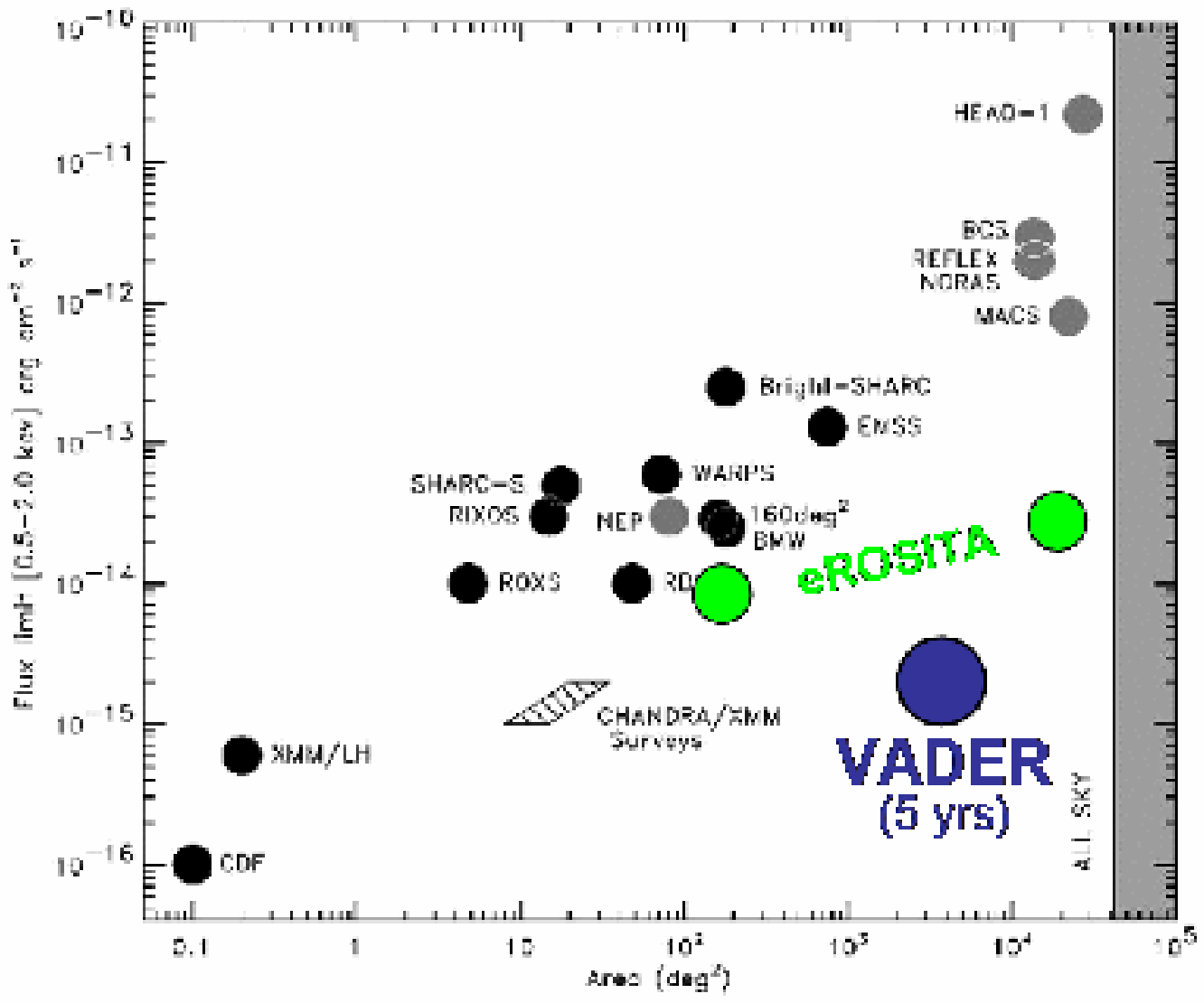}
\includegraphics[width=9.0cm]{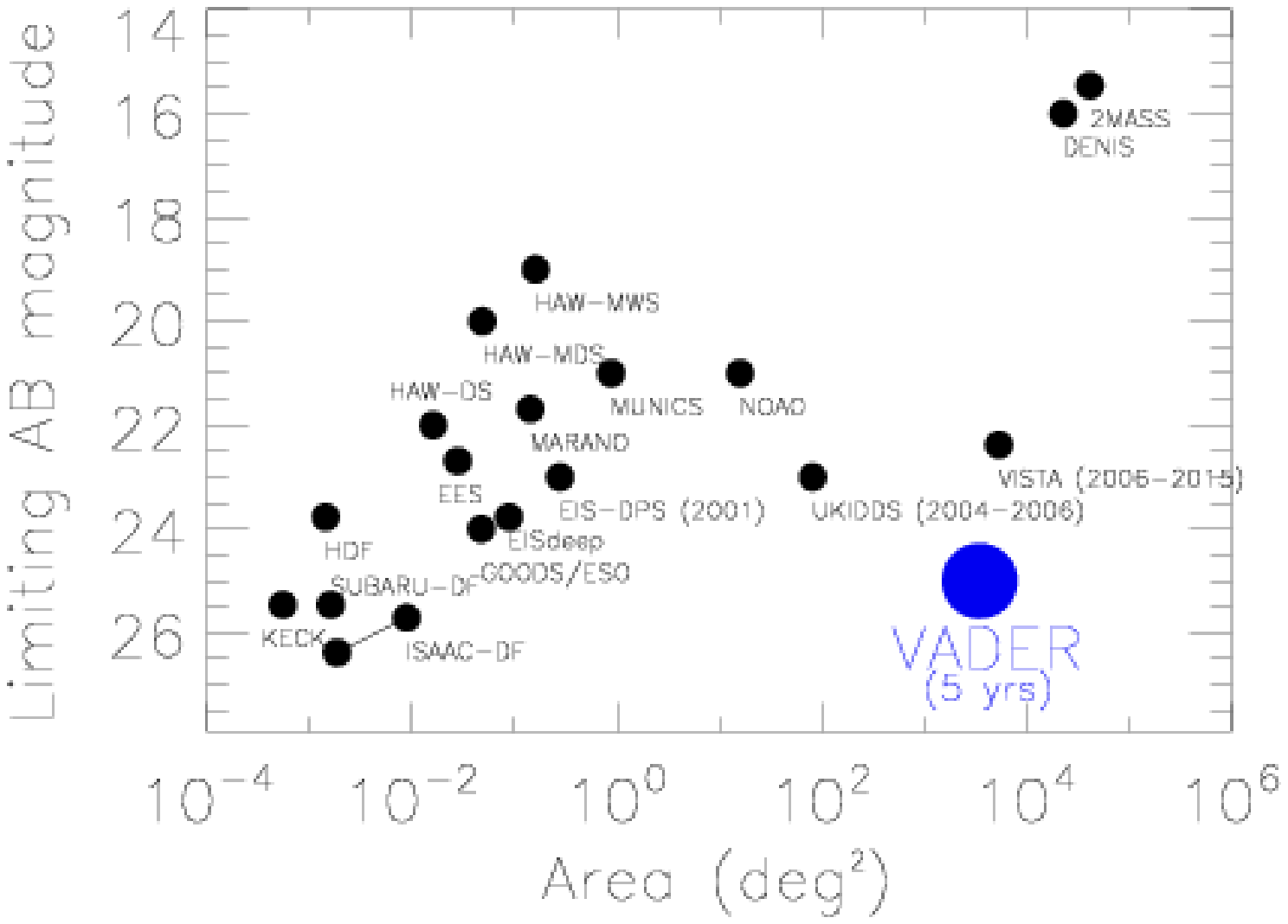}
\caption{VADER compared to other surveys. {\bf Left:} X-ray surveys. The VADER X-ray survey is located in the lower right corner of the flux limit vs. area plane and is unprecedented in terms of the combination of these two fundamental parameters. It will be more than an order of magnitude deeper than eROSITA and more than three orders of magnitude wider than the deep XMM-Newton surveys such as COSMOS. Original plot appeared in Ref.\cite{Rosati2002a} {\bf Right:} NIR surveys. The H band comparison shows a similar situation as in X-rays.}
\label{rf_fig_surveys}       
\end{figure}

VADER sensitivities have been calculated assuming net exposure times of 32\,ksec (8.9\,h), i.e.\,we take into account an average extra data loss time of 4\,ksec, which can be due to solar flare periods, readout overheads or detector malfunctioning.  
The limiting magnitudes for 10\,$\sigma$ detections in the optical/IR bands are listed in Tab.\,\ref{t_fluxlimits}. The X-ray flux limits for detections of extended and point-like sources are given in Tab.\,\ref{t_XrayPayload}. With these capabilities, VADER can detect and characterize all clusters of galaxies out to redshifts of $z$$\le$2 (mass limit M$_{lim}$$\sim1.5 \cdot 10^{14}$\,M\sun \ at $z$=2), detect galaxies up to $z$$\la$3, and study the $z$$\la$4 AGN population.
Figure\,\ref{rf_fig_surveys} compares the VADER sensitivities to other existing or planned X-ray and NIR surveys.


\begin{table}[b]
\begin{center}
\begin{tabular}{|c|c|c|c|c|}
\hline

 {\bf Band} &  {\bf Center} &   {\bf Limiting AB Magnitude} &  {\bf L$_*$ at $z$=2} \\
\hline\hline

u  & 0.35 $\mu$m   & 25.4 & 26.4 \\ 
g  & 0.48 $\mu$m   & 26.4 & 26.2 \\ 
r  & 0.63 $\mu$m   & 27.1 & 25.7 \\
z  & 0.90 $\mu$m   & 26.5 & 25.5 \\
J  & 1.2 $\mu$m    &  25.4 & 23.0 \\
H  & 1.6 $\mu$m    &  25.2 & 21.9 \\
K  & 2.2 $\mu$m    &  23.2 & 21.3 \\
L  & 3.6 $\mu$m    &  22.7 & 21.2 \\

\hline
\end{tabular}
\caption[Limiting magnitudes]{Limiting magnitudes for the VADER multi-wavelength survey. The table contains the band name, the central wavelength, the expected limiting AB magnitude for 10\,$\sigma$ detections, and the expected apparent magnitudes for a passively evolving L$_*$ galaxy at redshift $z$=2 for comparison.}
\label{t_fluxlimits}
\end{center}
\end{table}

\subsection{Science Data} 

Transmission constraints limit the downlinked data volume to about 1TB per orbit (see Sect.\,\ref{sect_comm}). While the data rates of the X-ray detectors are modest, the giga-pixel optical and infrared cameras (Sect.\,\ref{sec_payload}) produce raw data volumes that are beyond the transmitting capability.
In order to obtain a manageable data amount, the acquired optical/IR science data of the 2h observation blocks per pointing have to be combined to a single data set per camera, i.e. to about 30 full 2\,h data sets per orbit. This requires onboard co-addition of the individual camera exposures of typically 10-30\,min to single 2\,h master exposures. The expected progress of lossless data compression schemes could significantly increase the number of transmittable data sets. 

The main data reduction and processing is performed on ground. The full mission data volume of up to 1000\,TB requires a dedicated data center with sufficient processing power.
Multi-wavelength detection and classification tools for large data sets are currently under development\footnote{More details on the ClusterFinder of the German Astrophysical Virtual Observatory can be found here: \url{http://www.g-vo.org/clusterfinder}}
 and will be available in the near future.






\subsection{Expected Results}

The expected number of detected test objects and the resulting constraints on $w$ for the five main DE methods are listed in Tab.\,\ref{t_survey results}. With the given sensitivities, VADER will probe 50\,Gpc$^3$ of comoving volume with multi-wavelength coverage and is expected to detect 175,000 galaxy clusters\cite{Alexis2006a}, one billion galaxies\cite{Chen2002a}, and 5 million AGN\cite{Brandt2005a} during the nominal 5 year survey. To put this into perspective, compared to the currently $<$10 known galaxy clusters at redshifts beyond unity\cite{Mullis2005a}, VADER will find and characterize thousands of clusters at $z$$>$1 and hundreds of these sensitive cosmological probes at $z$$>$1.5.


The stated $w$ constraints in Tab.\,\ref{t_survey results} are conservative upper limits taken from the literature. Since one of the main science goals is `assumption free' DE tests, we do not want to overestimate the achievable accuracy of the individual methods. However, due to the sometimes orthogonal degeneracies of the different methods, the sub-percent precision regime for    
$\sigma_w$ is accessible for the combined analysis.
Quantitative constraints on the time variation of $w$ are harder to give, in particular in combination with the goal of using non-parametric tests. With the increased redshift leverage and the high statistical power, however, we expect that VADER's time variability measurement of $w$ will be unrivalled. 


The described data set will allow many  sensitive cosmological tests in addition to the described  prioritized list of methods. Just to name a few, the shape of the power spectra, the profile of dark matter halos, and the baryon fraction method with galaxy clusters could yield additional valuable information on DE. Furthermore,
the spin-off science potential of the VADER survey data set is substantial. Considering the science activities and results of recent surveys in X-rays (ROSAT) or optical (SDSS), this deep multi-wavelength survey will certainly trigger a multitude of new studies and breakthrough discoveries.





\begin{table}[tb]
\begin{center}
\begin{tabular}{|c|c|c|c|c|c|c|}
\hline

 {\bf ID} &  {\bf Method} &  {\bf \num \ of Test Objects (5yrs)} &  {\bf $\sigma_w$} &    {\bf Reference}  \\
\hline\hline

1 & BAO in galaxy cluster PS  & 175,000 galaxy clusters & $<5$\%  &   Angulo et al. (2005) \cite{Angulo2005a}   \\
2 & cosmic shear & 10$^9$ galaxies  & $<$5\%  &   Huterer et al. (2006) \cite{Huterer2006a}  \\
3 & BAO in galaxy PS  & 10$^9$ galaxies  & $<$5\%  & Blake et al. (2005) \cite{Blake2006a}    \\
4 & cluster number counts & 175,000 galaxy clusters & $<$10\%  &   Jahoda et al. (2002)\cite{DUETa}   \\
5 & BAO in AGN PS & $5 \cdot 10^6$ AGN  & $<10$\%  &    \\
\hline
 & all methods combined  &   & $\la$1\%  &    \\

\hline
\end{tabular}
\caption[Expected survey results]{Expected results and precision of the $w$ measurements of a 5 year  VADER survey. The survey will cover 3,500\,deg$^2$ and will probe a comoving volume of 50\,Gpc$^3$ for all DE tests ($z$$\le 2$). References are given for the $\sigma_w$ estimate. By combining all independent VADER measurements of $w$ with their different degeneracies, a sub-percent level precision can be reached. An extended 10 year survey could increase the survey area to 7,000\,deg$^2$, the comoving volume to 100\,Gpc$^3$, double the number of test objects, and thus allow an even more precise determination of $w$ and its time derivative.}

\label{t_survey results}
\end{center}
\end{table}


\subsection{Competitiveness}
Ground based experiments (KIDS, DES, APEX-SZ, HETDEX, WFMOS) typically aim for a 5-10\% error on $w$. The large surveys Pan-STARRS and LSST with their 20,000--30,000\,deg$^2$ sky coverage expect higher constraints. However, the lack of deep NIR coverage limits the redshift range and degrades the photo $z$ accuracy for DE tests using galaxies. Furthermore, the seeing limited resolution results in a significant degradation of the cosmic shear measurements, and some tests (see Tab\,\ref{t_methods}) rely on X-ray data and are therefore not feasible from the ground at all.
In comparison to other planned DE space missions, VADER provides the most complete and versatile survey data set for dark energy and general astrophysical studies. The X-ray survey mission eROSITA has a priori no redshift information and only reaches a survey depth which is 15 times shallower than VADER's. DUNE is a designated cosmic shear mission providing high resolution data only in a single optical band. SNAP's main science driver are distant supernovae (SNe) Ia, complemented by cosmic shear studies. The SNe  precision on $w$ will probably be limited by systematics and SNAP's possibilities for alternative DE tests and additional science are more restricted compared to VADER's.

\section{PAYLOAD}
\label{sec_payload} 
The payload consists of two XMM-Newton type X-ray telescopes in wide-field configuration and the 1.5m optical/IR telescope with its eight simultaneous operating cameras. In this section, we describe the different payload components in more detail.

\subsection{X-ray Telescopes}


The VADER X-ray telescopes are based on the XMM-Newton design, a pointing observatory with good angular resolution and the highest throughput currently available. The use of the well-calibrated XMM-Newton spare mirrors for a wide field survey mission has been studied before by the DUET\cite{DUETa} collaboration. Our design specifications are based on the measured spare mirror performance.

\begin{figure}[t]
\centering
\includegraphics[height=4.8cm]{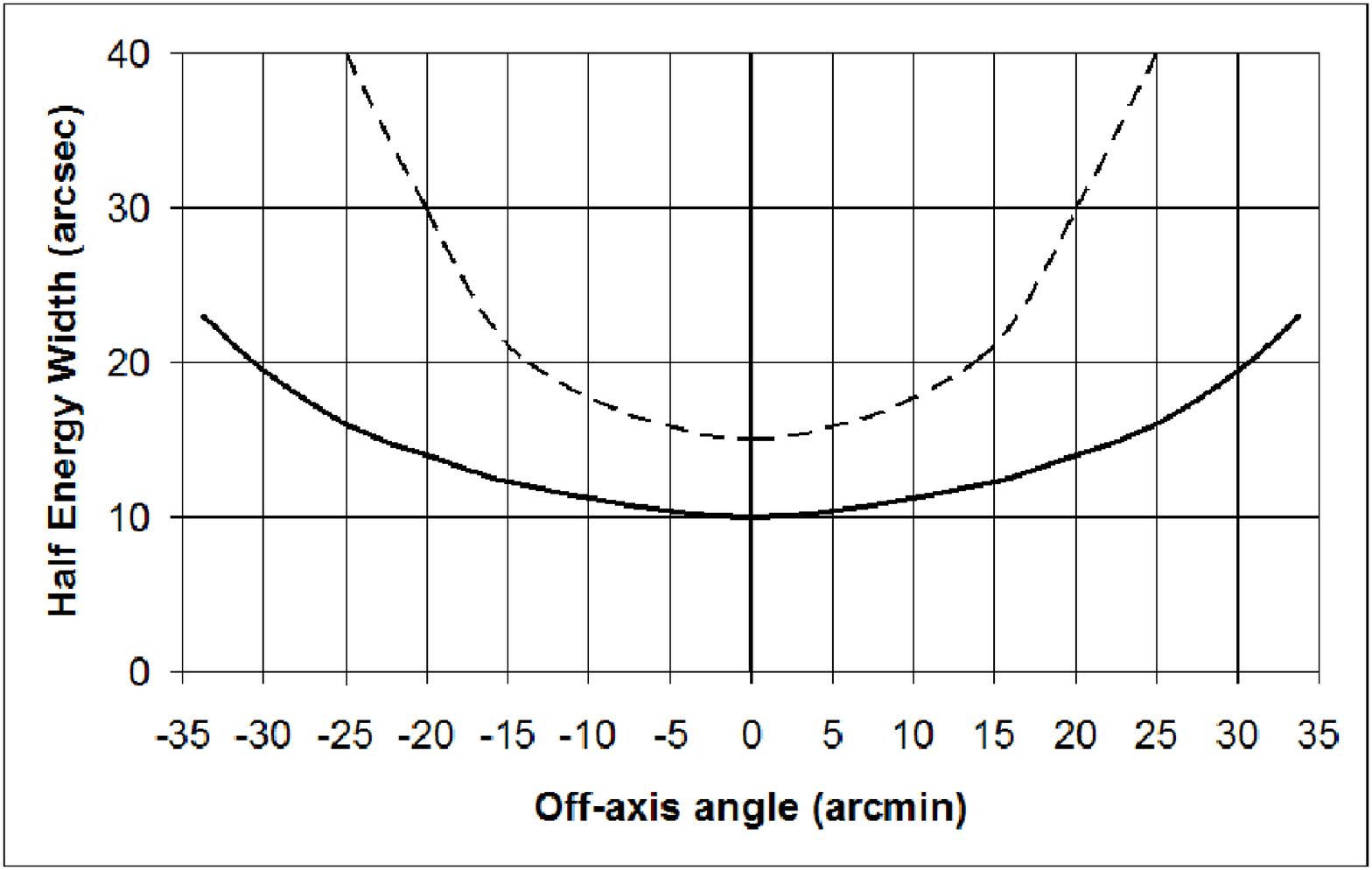}
\includegraphics[height=5.5cm]{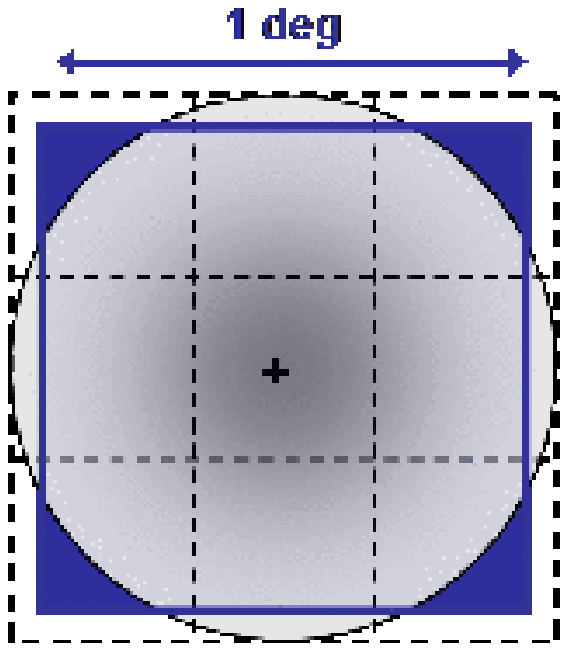}
\caption{{\bf Left:} Measured off-axis performance of the XMM-Newton spare mirror. The solid line indicates the half energy width (HEW) vs. off-axis angle performance of a detector configuration that follows the curved focal plane of the mirror as implemented in VADER, the dashed line shows the characteristics in the XMM-Newton configuration. The $\sim$20\,arcsec HEW over the VADER field of view is sufficient to resolve most galaxy clusters at any redshift. {\bf Right:} Matched field of views of the different telescopes. The 1\,deg$^2$ FoV of the optical/IR telescope is indicated by the inner solid square. The FoV of the X-ray telescopes is shown by the outer dashed lines, whereas the inner ones sketch the individual DEPFET detector arrays. The circle indicates the 1\,deg$^2$ sky area enclosed by an off-axis angle of 33.8\,arcmin from the optical axis (cross). The shading represents the increased vignetting of the X-ray mirrors towards the edge of the field.}
\label{rf_fig_xperformance}       
\end{figure}

The off-axis resolution performance of the XMM-Newton mirrors can be significantly improved, if the detectors follow the curved focal plane of the grazing incident mirrors. This resolution improvement is shown in the left panel of Fig.\,\ref{rf_fig_xperformance} resulting in a sufficiently good half energy width (HEW) of $<$23\,arcsec over a 1 square degree field of view (displayed in the right panel of Fig.\,\ref{rf_fig_xperformance}). Although the vignetting effects of the mirrors are significant with increasing off-axis angle, the grasp --- the product of the effective area and the solid angle $A\!\cdot\!\Omega$ --- is still very good. By using two X-ray telescope systems, VADER reaches a total effective grasp (at 1.5\,keV) of 1120\,cm$^2$deg$^2$, surpassing the eROSITA X-ray survey satellite by more than 50\%. The main characteristics of the VADER X-ray telescopes and detectors are summarized in Tab.\,\ref{t_XrayPayload}.


The two wide-field X-ray imaging cameras are based on state-of-the-art
DEPFET active pixel sensors\cite{Fischer2000a} 
in which  the amplifying transistor structure is contained in the pixel cell itself.
Due to the good energy and spatial resolution, the high quantum efficiency, the fast readout and low noise performance, and their low power consumption, DEPFETs are the technology of choice for the VADER mission.



\begin{figure}[t]
\centering
\includegraphics[width=12.5cm]{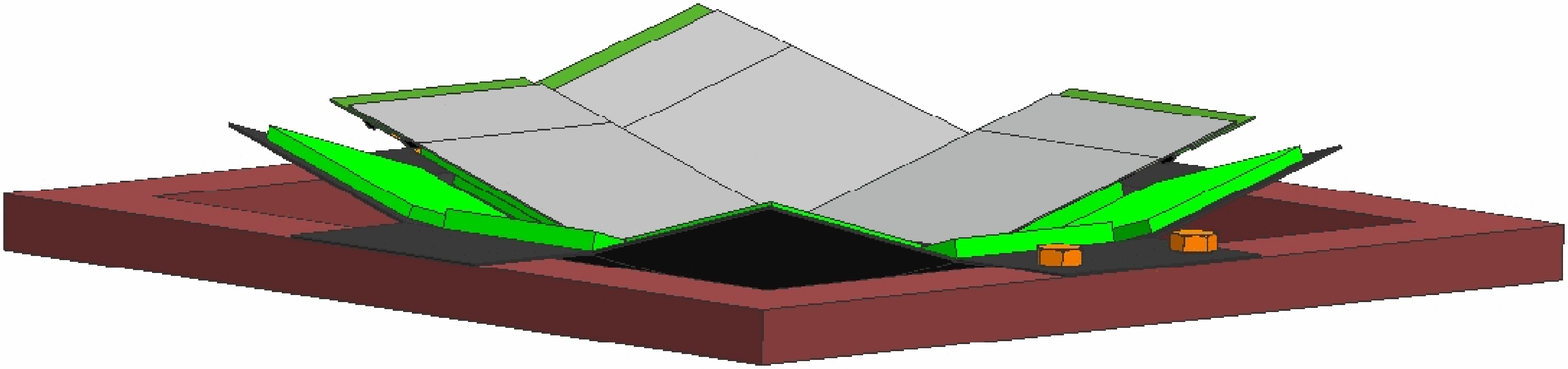}
\includegraphics[width=6.5cm]{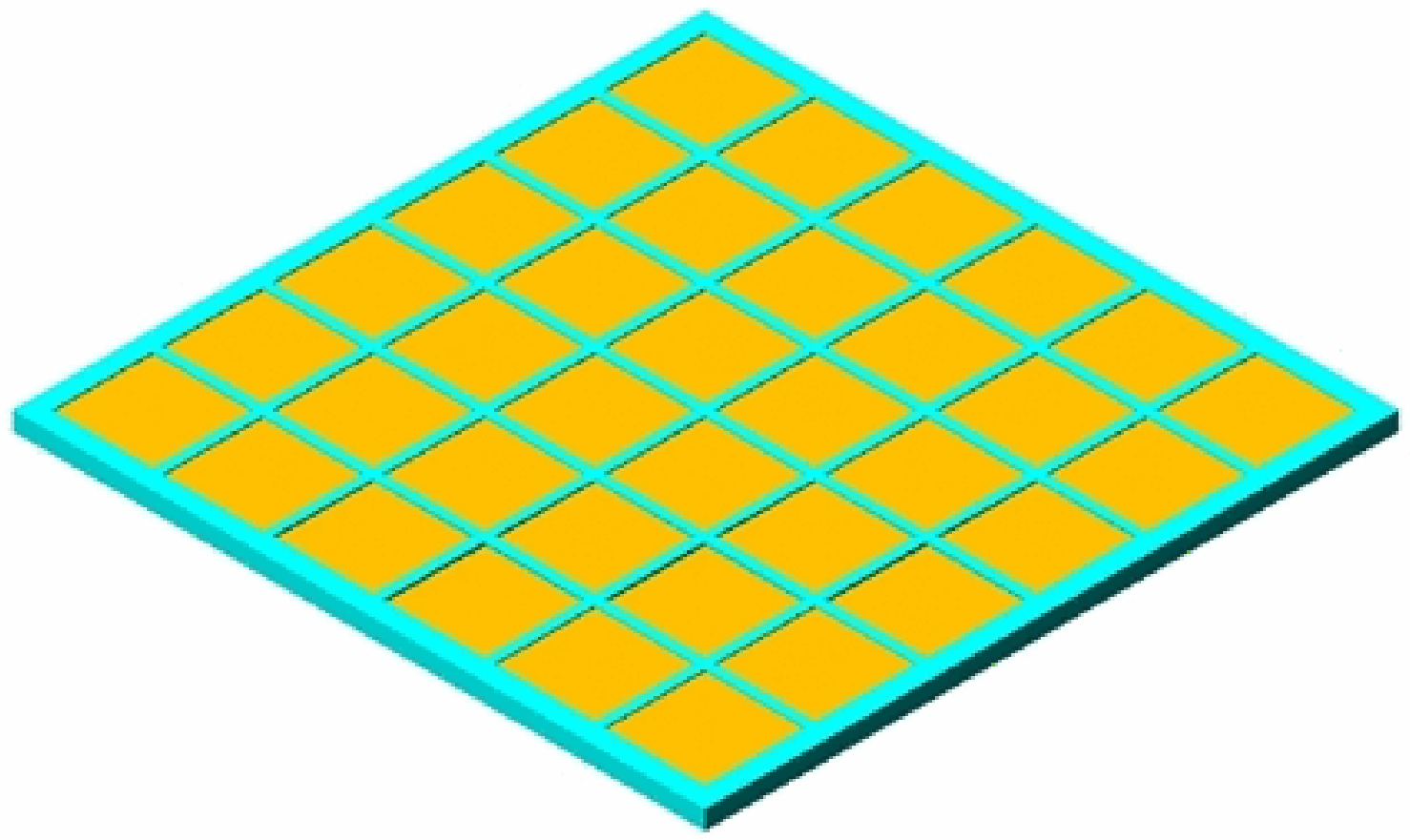}
\includegraphics[width=6.5cm]{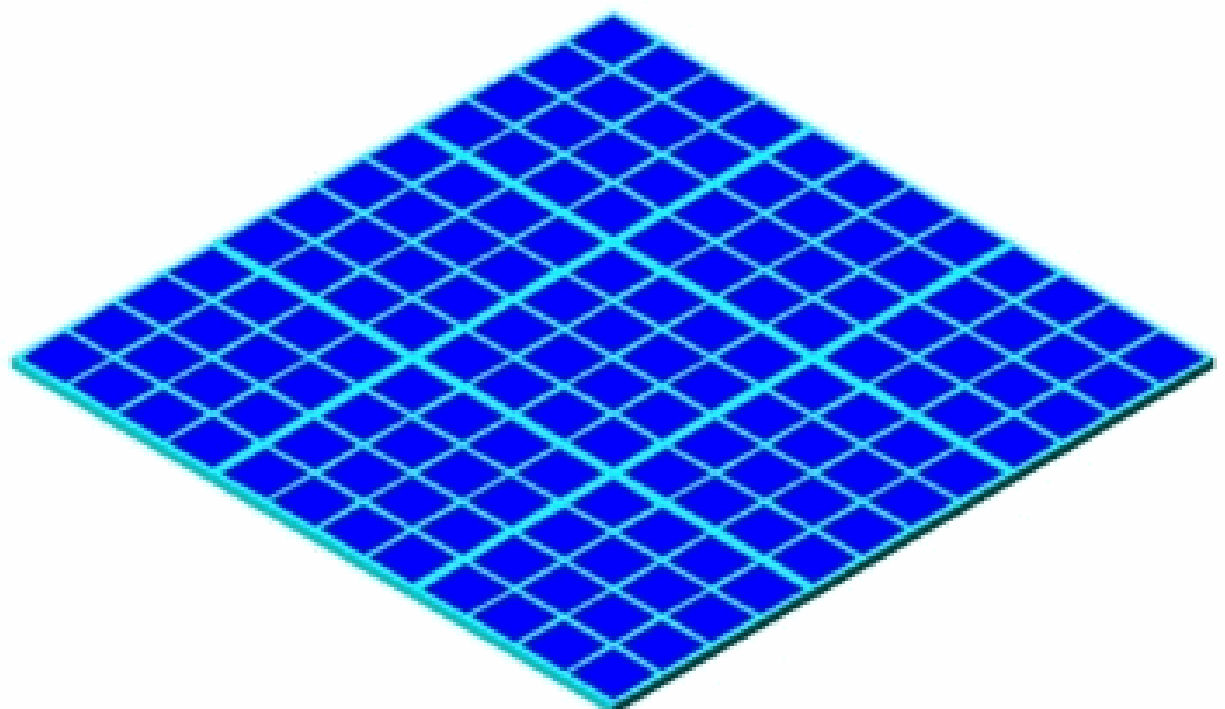}
\caption{Schematic views of the VADER detector systems. {\bf Top:} Scheme of the wide-field X-ray detector system in the curved focal plane configuration. The camera consists of 9 512\,$\times$\,512 DEPFET arrays. The outer arrays are tilted to optimally follow the parabolically curved focal plane. {\bf Bottom:} Layout of the optical and IR detector planes. The z-band camera ({\it left}) consist of 36 10k\,$\times$\,10k CCDs. The J, H, and K band cameras ({\it right}) use 144 2k\,$\times$\,2k NIR arrays. The u, g, r, and L-band cameras are scaled down versions of these two layouts.}
\label{rffig_xdetector}       
\end{figure}

A great advantage of the pixel readout compared to CCD devices is the high integration-to-readout-time ratio, thus optimizing the effective exposure time while preserving the full spatial information (no out-of-time-events). The 3\,$\times$\,3 detector mosaic (see Fig.\,\ref{rffig_xdetector}) is based on 512\,$\times$\,512 pixel arrays with dimensions 5.1\,$\times$\,5.1cm and pixel sizes of 100\,$\mu$m.
In order to improve the off-axis resolution performance (Fig.\,\ref{rf_fig_xperformance}) and extend the FoV  to one square degree, the outer detectors are tilted to  follow the parabolic focal plane as obtained from calibration measurements. 
To further reduce the particle induced background  from surrounding structures, the detector box is built from carbon fiber and honeycomb structures.

\begin{table}[t]
\begin{center}
\begin{tabular}{|c|c||c|c|}
\hline

number of mirror systems & 2 & detector type   & DEPFET   \\   
focal length & 750\,cm & pixel size   & 100\,$\mu$m   \\   
diameter of mirror & 70\,cm & detector size   &  1.5k\,$\times$\,1.5k  \\   
total weight of mirrors & 900\,kg & read out time   &  2\,ms   \\
on-axis effective area (1.5\,keV)  &  2800\,cm$^2$   & energy resolution (1.5\,keV)  &   60\,eV \\
average vignetting factor   &  0.4  & net exposure time per field & 32\,ksec \\
field of view   &  1\,deg$^2$  & extended sources flux limit  & $2 \cdot 10^{-15}$\,erg\,s$^{-1}$\,cm$^{-2}$ \\
total grasp $A\!\cdot\!\Omega$ & 1120\,cm$^2$\,deg$^2$ & cluster surface density at f$_{lim}$ & 50\,deg$^{-2}$\\
half energy width   &  $<$23\,arcsec  & point sources flux limit  & $1 \cdot 10^{-15}$\,erg\,s$^{-1}$\,cm$^{-2}$ \\
energy range & 0.1-10\,keV & AGN surface density at f$_{lim}$ & 1500\,deg$^{-2}$\\

\hline
\end{tabular}
\caption[Characteristics of X-ray Payload]{Characteristics of the X-ray mirrors ({\it left}) and the wide-field X-ray detectors with expected sensitivities ({\it right}). The galaxy cluster and AGN surface densities at the flux limit are taken from Refs. \cite{Alexis2006a} \& \cite{Brandt2005a} .}

\label{t_XrayPayload}
\end{center}
\end{table}

\subsection{Optical/IR Telescope} 


VADER's 1.5\,m optical/IR telescope is based on a 2-mirror modified Ritchey-Chretien design with a corrected wide field of view of 1 square degree. The two hyperbolic mirrors form an f/16.6 system\footnote{There are currently also designs with different f ratios under consideration. The final parameters of the optical system with the best image quality have to be determined during the phase A study using detailed ray-tracing simulations.}, with the effective focal ratio of the individual cameras being adjusted by means of focal reducers and enlargers. The M1 and M2 mirrors are fabricated with the lightweight honeycomb sandwich mirror technology using a porous structure core of silicon carbide (SiC) with a dense SiC coat. 
The resulting very high specific stiffness and the low sensitivity to thermal gradients, in combination with the low weight, is ideal for a space based mission. Silver coating of the mirrors provides a reflectivity of $>$90\% from the u to the L band.  

\begin{figure}[t]
\centering
\includegraphics[width=12.0cm]{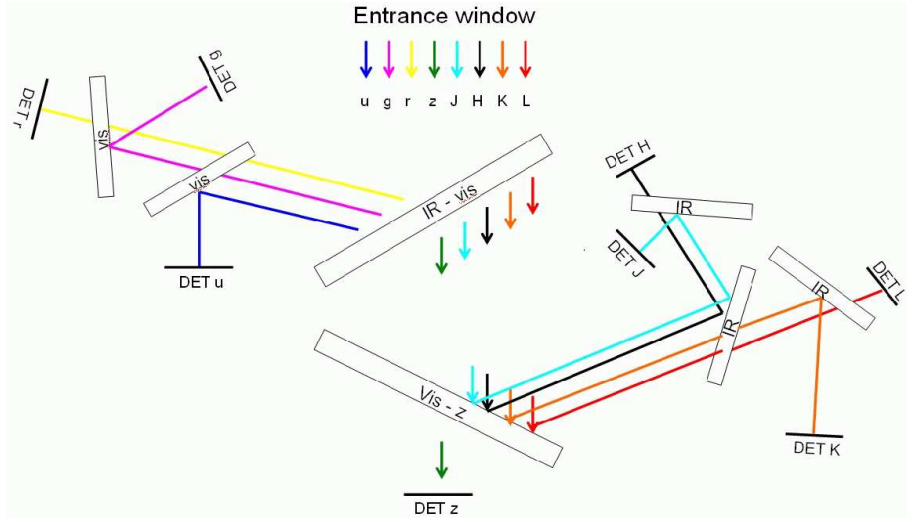}
\caption{Beam splitting scheme via dichroics and camera configuration. The incoming beam is separated into eight parts by means of seven dichroic beam splitters. The z-band camera, which has the highest optical quality requirement, is located in the Cassegrain focus of the telescope, the beams for the other seven cameras are redirected.}
\label{rffig_beamsplitting}       
\end{figure}

The main challenge of the optical design is to achieve high optical quality with diffraction limited resolution (in z, K, L) over the 1 square degree FoV in combination with an eight channel system.
The beam splitting scheme for the eight bands is shown in Fig.\,\ref{rffig_beamsplitting}. For a small FoV a similar strategy has been successfully applied and tested by the ground-based GROND instrument\footnote{Details on the GROND instrument can be found here: \url{www.mpe.mpg.de/~jcg/GROND}}. The highest resolution and optical quality demands are required for the z-band camera, which is therefore located in the Cassegrain focus with only two dichroic transmissions of the beam required. 
The remaining IR and optical stages are
branched off from the main beam and redirected to a different satellite compartment where the beam is further separated. A total of seven dichroics is necessary in order to obtain the eight bands.

The main characteristics of VADER's eight optical and IR camera systems are summarized in Tab.\,\ref{t_opt_cam}.  
The z, K, and L-band cameras (see bottom panel of Fig.\,\ref{rffig_xdetector}) feature critical PSF sampling for a diffraction limited resolution. The u, g, r, J and H-band detectors are undersampled but still have a resolution of $\la$0.15\arcsec, much better than seeing limited ground-based instruments. All eight detectors together comprise 8.7\,billion pixels. 
The four optical detector systems (ugrz) make use of newly developed, back side thinned 10k\,$\times$\,10k\footnote{R.A. Bredthauer and M. Lesser, {\it Siliconus Maximus}, SDW2005 Scientific Detector Workshop, June19-25 2005, Taormina, Italy} CCD chips with fast 10s readout, high quantum efficiency out to 1$\mu$m, and a pixel size of 9$\mu$m. The J, H, K, and L infrared cameras are based on HAWAII-2\footnote{www.rockwellscientific.com/imaging/hawaii2rg.html} 2k\,$\times$\,2k HgCdTe arrays, which will be available as 8k\,$\times$\,8k mosaic units in the near future. The band gap of the HgCdTe detector material can be adjusted to optimally match the wavelength range of the individual IR cameras. 
Each of the eight camera systems is equipped with a non-movable filter (see Tab.\,\ref{t_opt_cam}) and a multi-lens wide field corrector to accomplish high optical quality.

\begin{table}[tb]
\begin{center}
\begin{tabular}{|c|c|c|c|c|c|c|c|c|c|}
\hline

 {\bf Band} &  {\bf Center}  &   {\bf Width}  &  {\bf Layout} & {\bf Pix} & {\bf \num Pixels} &  {\bf Pix. Scale} &  {\bf Diff. Lim.} &  {\bf Size} & {\bf f Ratio} \\

 &    [$\mu$m] &    [$\mu$m] &  & [$\mu$m] &  [10$^6$\,pix] &  [\arcsec / pix] &  &  [cm]  &\\

\hline\hline

u  & 0.35  & 0.05  & 9 10k\,$\times$\,10k & 9 & 944 & 0.13  &  no &  28 & 9.7 \\
g  & 0.48  & 0.10  & 9 10k\,$\times$\,10k & 9 & 944 & 0.13  &  no & 28  & 9.7 \\
r  & 0.63  & 0.10  & 9 10k\,$\times$\,10k & 9 & 944 & 0.13  &  no & 28 & 9.7  \\
z  & 0.90  & 0.12 & 36 10k\,$\times$\,10k  & 9 & 3775 & 0.064  &  yes & 55  & 19.4 \\

\hline

J  & 1.2  & 0.26  & 144 2k\,$\times$\,2k  & 18 & 604 & 0.15  &  no & 44  & 16.6 \\
H  & 1.6  & 0.29  & 144 2k\,$\times$\,2k   & 18 & 604 & 0.15 & no & 44  & 16.6 \\  
K  & 2.2  & 0.41  & 144 2k\,$\times$\,2k   & 18 & 604 & 0.15  & yes  &  44  & 16.6 \\
L  & 3.6  & 0.57  & 64 2k\,$\times$\,2k  & 18 & 268 & 0.25  & yes  & 29  & 10.2 \\

\hline
\end{tabular}
\caption[Overview optica/IR cameras]{Main characteristics of VADER's eight optical and IR cameras. 
From left to right the table contains the band name, the central wavelength, the band width, the camera layout, the pixel size, the number of pixels per camera, the pixel scale, the diffraction limit flag, the camera diameter size, and the effective f ratio. The upper four lines list the optical cameras with CCD technology and an operating temperature of $\sim$150\,K, the lower four rows show the IR cameras using NIR array technology at a temperature of  $\sim$35\,K. The effective f ratio is adjusted to the camera by means of focal reducers/enlargers.  The combined number of pixels of all cameras is 8.68\,Gpix.} 

\label{t_opt_cam}
\end{center}
\end{table}

\subsection{Design Alternatives} 
For the X-ray telescopes, the current VADER design is based on the XMM-Newton mirror technology, which was not optimized for wide field imaging. Although this design was shown to fulfill all mission specifications and has the advantage to be readily available at low cost, the off-axis vignetting and resolution performance could be significantly improved with some R\&D efforts.

In principle, the first three optical bands u, g, and r could be imaged using ground-based telescopes. Taking the cameras for these bands out, VADER's data volume would be reduced by $\sim$30\%, the weight by $\sim$10\%, and the costs  by 5-10\% (not accounting the costs for ground-based observations). Since VADER images the bands simultaneously and the overall size and weight are not considerably affected, the inclusion of optical u, g and r bands has significant advantages for the data quality and the related science compared to ground-based observations for several reasons: (i) the homogeneous high data quality with very stable conditions, (ii) the resolution is still more than twice as good and prevents source confusion, (iii) the cross-calibration of the bands is much easier and more accurate, (iv) the photometry and thus the photometric redshifts are more accurate, and (v) the full time domain ability is maintained.


\section{SPACECRAFT} 


\subsection{Global Properties and Satellite Structure} 

\begin{figure}[t]
\centering
\includegraphics[height=5.3cm]{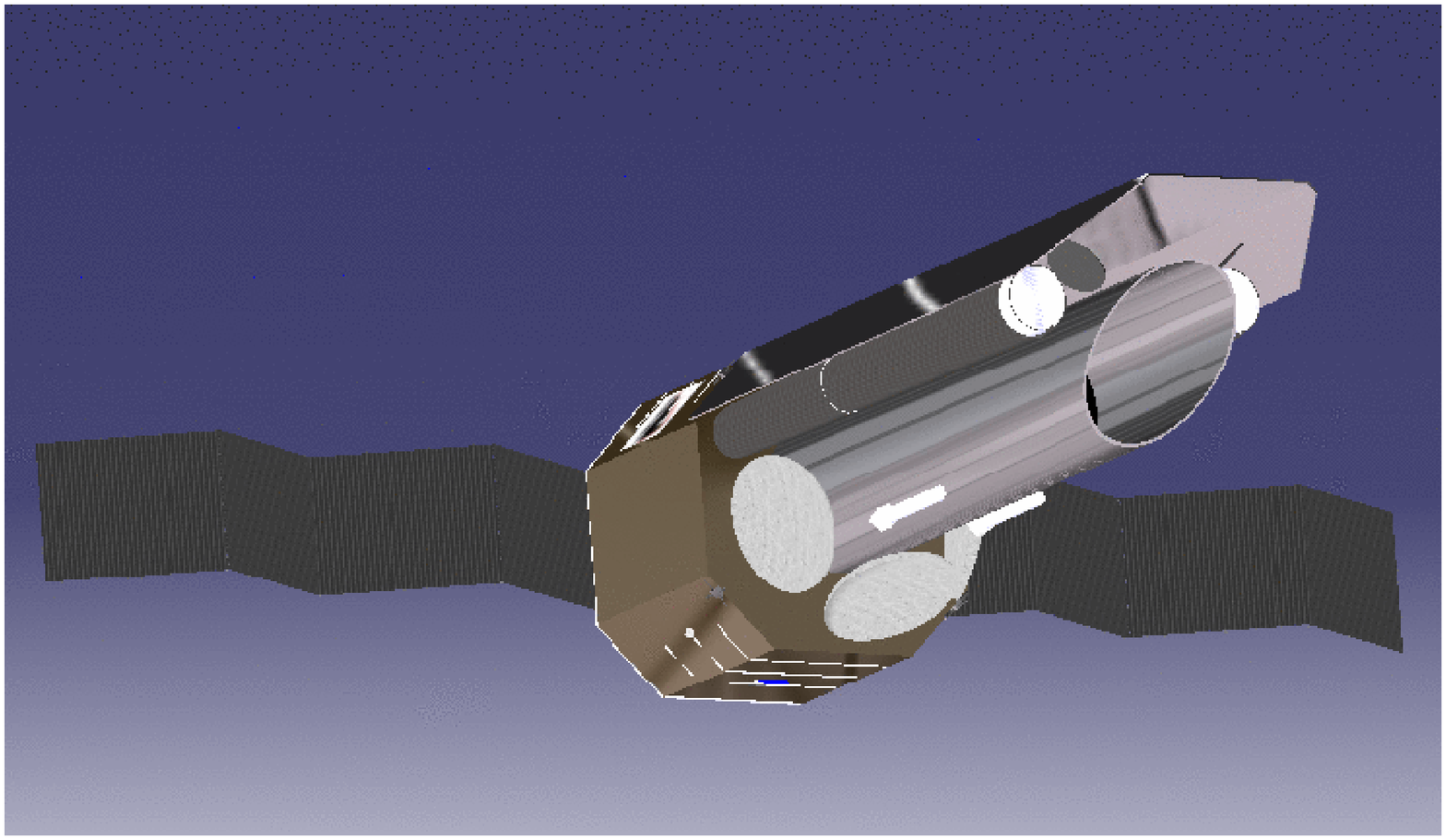}
\includegraphics[height=5.3cm]{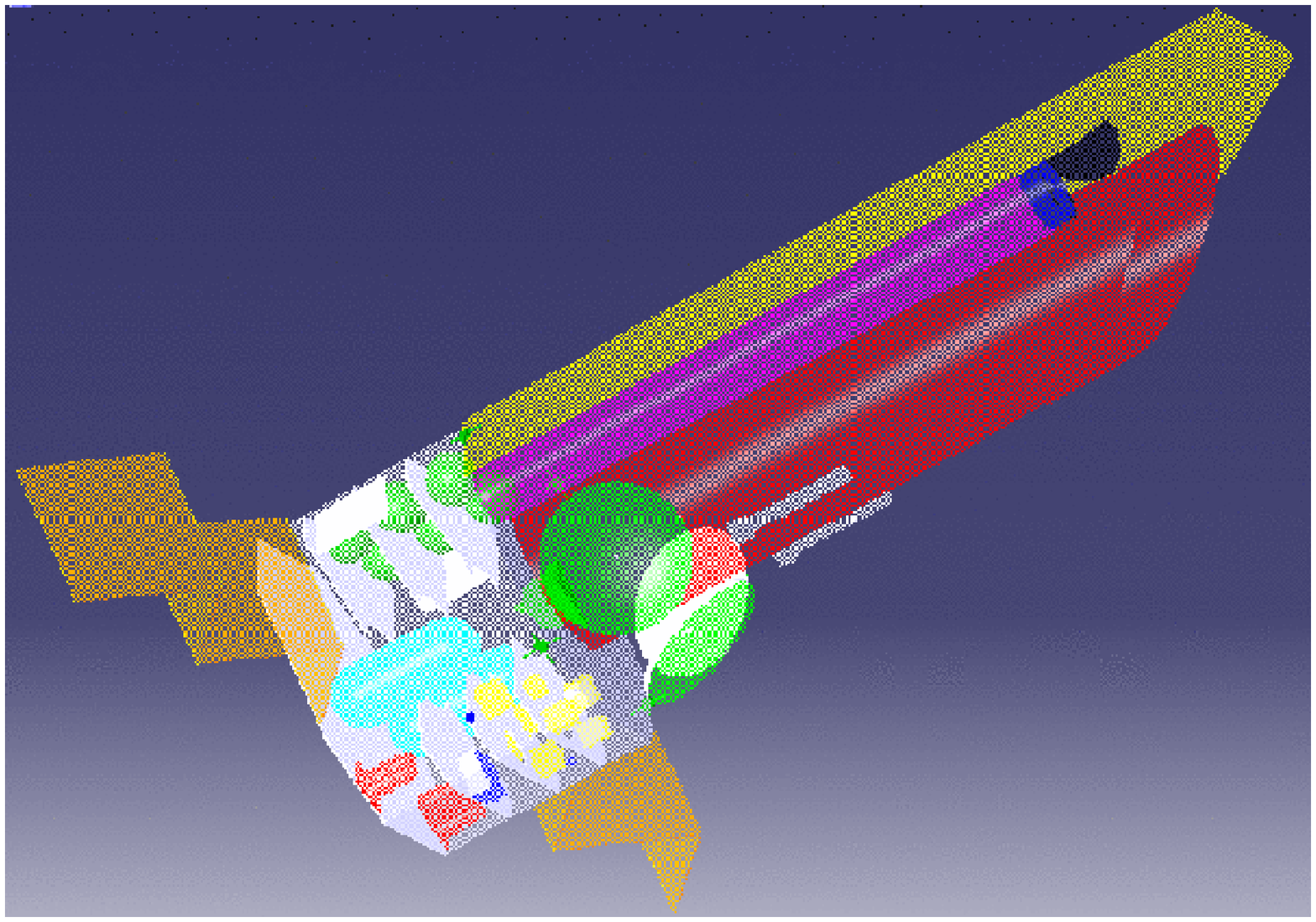}
\caption{Scheme of the VADER spacecraft {\it (left)} and in transparent view {\it (right)}. }
\label{rffig_spacecraft}       
\end{figure}

The VADER satellite is shown in Fig.\,\ref{rffig_spacecraft}, with basic key figures on the mass and power budget given in Tab.\,\ref{t_powermass}.
The spacecraft bus is made of a honeycomb structure, with a central cylinder for added rigidity. The bus has an 8-edge layout, with two internal walls, each going from one outer edge to the other, interrupted only by the central cylinder. These four walls separate the bus into 4 thermally separated rooms: 2 for instruments and 2 for electrical systems (CPU, heating). The solar panels with a total area of 15.1\,m$^2$ are deployable and can rotate around one axis, an additional sunshield has the dimension 8\,$\times$\,4m and a weight of 10\,kg. As standard interface to the launcher an aluminium adapter will be used. The total mass of the satellite structure will be around 900\,kg.

For the Attitude Determination and Control System (ADCS) two star sensors are used. They provide good attitude accuracy with reasonable weight and power consumption. A combination of external and internal torquers is implemented. Whereas the external gas jets, fuelled by six tanks, have the advantage of being insensitive to altitude and can torque about any axis, the internal torquers consisting of 
four control moment gyroscopes in a tetrahedral arrangement are suitable for three-axis control and for providing momentum bias.

\subsection{Thermal Control and Power Subsystems} 
The temperature control system consists of six parts. A Mylar sheet on a honeycomb structure makes up the 8\,$\times$\,4m solar shield that protects the whole spacecraft, whereas the telescopes are  wrapped in additional layers of Mylar. The passive cooling system consisting of a 2\,m$^2$ radiator further decreases the temperature. The IR detectors are cooled by 6 Stirling closed-cycle coolers to a temperature of approximately 35\,K, while their heat output is  radiated away by a second 1.4\,m$^2$ radiator. The X-ray mirrors are operating at higher temperatures and are maintained by two 40\,W heaters. 
The different spacecraft compartment rooms are thermally isolated from each other. During eclipse phases of a maximal duration of 1 hour, heaters are used to maintain the necessary temperatures. 
Four solar panels with deployable wings and rotation ability around one axis are the primary power source. The cells are made from Gallium Arsenide Triple Junction Solar Cells with an efficiency of 23.8\%
and a  maximum power output of 4.2\,kW. As secondary power source Nickel-Hydrogen batteries are used.


\subsection{Communications} 
\label{sect_comm}

To provide an optimum communications subsystem, VADER is equipped with a 0.5 m transmitter antenna using the Ka-band at 22\,GHz for downlink.  During the telecommunications period of approximately 10\,h a downlink to the ground station in Kourou (15\,m antenna) and in Malindi (15\,m antenna) is established. To ensure a total amount of 1 Terabyte data volume per orbit, a high data rate downlink of 300\,Mbit/s is necessary. Hence, high-precision antenna-pointing mechanisms have to be realized, both at the transmitter of the satellite and of the ground station receivers, to implement a fast and precise tracking with sufficiently low pointing losses. One Travelling Wave Tube Amplifier (TWTA) with a transmitting output power of 10\,W is used at the satellite and a second one for redundancy. The link budget and the access times are computed for a maximum range of 60,000\,km between the satellite and the considered ground station, which guarantees a sufficient margin. Both ground stations are connected via the recently modernized ESATRACK Network and can therefore operate together to achieve a total access time of at least 10\,h (time used for handover and tracking already regarded). The first 5\,h of the telecommunication period will be downlinked to the ground station in Kourou, after the handover the ground station in Malindi will continue to receive the stored data. The station in Maspalomas (15\,m antenna) can serve as an alternative at times when the link to the other ground stations is weak. This ground station can provide a total access time of at least 7\,h. 
The uplink is done in the X-Band (8\,Ghz) using a separate uplink antenna (low-gain, helix antenna) on the satellite. For telecommand a maximum data rate of 50\,kbit/s is sufficient to keep in contact with the satellite independent of its current altitude.\\
For emergency purposes two omnidirectional antennas are mounted on the spacecraft, so that contact to Earth can be guaranteed independent of orientation. An emergency link with a data rate of 1-2\,kbit/s is considered. \\




\begin{table}[t]
\begin{center}
\begin{tabular}{|c|c|c|}
\hline

 {\bf Subsystem} &  {\bf Mass [kg]}  &   {\bf Power Supply [W]}   \\

\hline\hline

 ADCS       & 1070  & 520  \\
 Structure   & 960  &  0 \\
 Power   & 250  &  250 \\
 Payload   & 1810  & 820  \\
 Communication    & 20  &  110 \\
 Thermal   & 120  & 530  \\
 Data   & 230  &  230 \\       
 
 \hline
 Total   &  4,460 & 2460  \\     
 {\bf Total with 20\% margin} &  {\bf  5,350}  &   {\bf 2,950}   \\ 

\hline
\end{tabular}
\caption[Satellite Subsystems]{Mass and power budget of the satellite subsystems. }

\label{t_powermass}
\end{center}
\end{table}

\subsection{Launch and Operations} 

The launch will take place from Kourou in French Guiana (5\degr 18\arcmin\, N, 52\degr 48\arcmin\, W) using an Ariane\,5 ECA upper stage rocket. The Ariane\,5 will send the satellite into its  highly elliptical 72\,hour orbit. With a satellite length of 10\,m and a total mass of about 5\,tons, the use of a powerful rocket such as Ariane\,5 is required. 
VADER will be operated and controlled from the three ground segments  Missions Operations Center (MOC), Science Operations Center (SOC), and the Instrument Control Center (ICC).
VADER could be realized within the scope of a typical ESA cornerstone or a medium sized NASA observatory mission with a launch date foreseen around the year 2020.

\section{CONCLUSIONS} 

 

We have presented  a satellite mission concept with which the dark energy equation of state parameter $w$ and its time variation can be constrained with unprecedented precision. VADER will allow detailed studies using different cosmological probes and independent DE tests with low systematics out to redshifts of 2 and beyond.

The special mission strengths are (i) the cross-check capabilities of several independent DE methods, (ii) the high redshift leverage, (iii) the high sensitivity multi-wavelength coverage, and (iv) the versatility of the survey data set for numerous additional astrophysical studies.
Whereas the X-ray part of the mission bears only low technological risk, the innovative multiplexing capability of the wide-field optical/IR telescope system will require a challenging optical design. The main R\&D activities for VADER will thus have to focus on the different optical components, the wide-field detector technology, and an optimized thermal structure. However, compared to other planned missions (e.g. JWST, LISA) the overall technological risks of VADER are quite moderate.   

Unveiling the very nature of the dominant dark energy component in the Universe will be VADER's main challenging task. 
Considering the fundamental importance and the observational difficulty, the scope of the proposed mission concept poses an adequate answer to this pressing question.
Even though the notion of dark energy might be replaced in the theories at some point, VADER's contribution to a profound understanding of the Universe will remain and could guide the way to new insights in fundamental physics.




\acknowledgments     
We would like to thank our additional team members Peter Asaro, Christina Henriksen, Claire Juramy, Lisa Kaltenegger, Luis Alberto Martinez Vaquero, Halvor Midthaug, Lavinia Nati, and Georg Zwettler for their support in developing this project. We also thank 
Peter Bitzenberger, J\"urgen Krail, Anja von der Linden, Christian Clemens, Mario Schweitzer, Gabriel Pratt, and Hans B\"ohringer for sharing their expertise and 
 helpful comments and discussions. Furthermore, we are very thankful to the organizers, lecturers, and tutors of the 2006 summer school in Alpbach, where this work originated, and to  ESA and all national space agencies and foundations, in particular ASA, DLR, SRON and  LKBF, for their financial support. 

\newpage
 



\bibliography{rf_report}   
\bibliographystyle{spiebib}   

\end{document}